\documentclass[twocolumn,superscriptaddress,nobibnotes,nofootinbib,floatfix,aps,prl,citeautoscript,reprint]{revtex4-1}

\usepackage{graphicx}
\usepackage{epsfig}
\usepackage{subfigure}
\usepackage{color}
\usepackage{latexsym}
\usepackage{amsmath,bm}
\usepackage{dcolumn}           
\usepackage{natbib}
\newcommand{\etal}{\textit{et al.~}}
\newcommand{\ub}{Department of Physics, Universit\"{a}t Basel,
Klingelbergstr. 82, 4056 Basel, Switzerland}

\newcommand{\lpmcn}{Universit\'e de Lyon, F-69000 Lyon, France and 
LPMCN, CNRS, UMR 5586, Universit\'e Lyon 1, F-69622 Villeurbanne, France}
\newcommand{\lsi}{Laboratoire des Solides Irradi\'es and ETSF, \'Ecole Polytechnique, 
CNRS, CEA-DSM, 91128 Palaiseau, France}

\begin{document}

\title{Prediction of a novel monoclinic carbon allotrope}

\author{Maximilian Amsler}
\affiliation{\ub}
\author{Jos\'e A. Flores-Livas}
\affiliation{\lpmcn}
\author{Silvana Botti}
\affiliation{\lsi}
\affiliation{\lpmcn}
\author{Miguel A.L. Marques}
\email{miguel.marques@univ-lyon1.fr}
\affiliation{\lpmcn}
\author{Stefan Goedecker}
\email{stefan.goedecker@unibas.ch}
\affiliation{\ub}

\date{\today}

\begin{abstract}
  A novel allotrope of carbon with $P2/m$ symmetry was identified during an \emph{ab-initio} minima-hopping structural search which we call $M10$-carbon. This structure is predicted to be more stable than graphite at pressures above 14.4~GPa and consists purely of $sp^3$ bonds. It has a high bulk modulus and is almost as hard as diamond. A comparison of the simulated X-ray diffraction pattern shows a good agreement with experimental results from cold compressed graphite.
\end{abstract}

\maketitle
Graphite and diamond are the thermodynamically most stable forms of carbon at ambient conditions. However, carbon can be found in a vast structural variety due to its flexibility to form $sp$-, $sp^2$- and $sp^3$- hybridized bonds: hexagonal diamond, nano-diamond, carbon-foams, fullerenes and nanowires are just some examples of known carbon allotropes. Experimental evidences of a novel super-hard carbon phase have been reported in literature when graphite is compressed at room temperatures. Changes in resistivity~\cite{bundy_hexagonal_1967}, optical transmittance~\cite{goncharov_graphite_1989,hanfland_graphite_1989}, optical reflectivity~\cite{utsumi_light-transparent_1991}, X-ray diffraction (XRD) patterns~\cite{zhao_x-ray_1989,yagi_high-pressure_1992,mao_bonding_2003} and in the raman spectra~\cite{amsler_crystal_2012} indicate a phase transition in the range of 10 to 25\,GPa. Recently, several candidate structures have been proposed to match these experimental observations, such as $M$-carbon~\cite{li_superhard_2009}, bct-C$_4$-carbon~\cite{umemoto_body-centered_2010}, $W$-carbon~\cite{wang_low-temperature_2011} and $Z$-carbon~\cite{amsler_crystal_2012,zhao_novel_2011}. Although $Z$-carbon is thermodynamically the most promising structure, a final and conclusive determination has not yet been possible.

In this article we report on a novel carbon allotrope discovered with the recently developed minima hopping crystal structure prediction method (MHM)~\cite{goedecker_minima_2004,amsler_crystal_2010}. The MHM is capable to predict the most stable and metastable structures given solely the chemical composition of a system. Short molecular dynamics simulations are used to escape from local minima, and local geometry relaxations are performed to identify stable configurations. High efficiency of the escape step is ensured by aligning the initial molecular dynamics velocities along soft mode direction, and revisiting already explored regions of the potential energy surface is avoided by a feedback mechanism. The minima hopping method has been successfully used in a wide range of applications~\cite{amsler_crystal_2012,hellmann_2007,roy_2009,bao_2009,willand_2010,de_energy_2011}. 

During our MHM simulations, the energies and the Hellman-Feynman forces were evaluated at the density functional theory (DFT) level within the local density approximation (LDA), and the all-electron projector-augmented wave method was employed as implemented in the {\sc abinit} code~\cite{gonze_brief_2005,bottin_large-scale_2008}. The most promising structures were further relaxed using norm-conserving Hartwigsen-Goedecker-Hutter (HGH) pseudopotentials~\cite{hartwigsen_relativistic_1998}. The total energy was converged within less than 1~meV per atom by a plane wave cut-off energy of 2100\,eV and well converged Mankhorst-Pack $k$-point meshes. We reconfirmed the energy ordering with two other exchange-correlation functionals within the generalized gradient approximation, namely PBE~\cite{perdew_generalized_1996} and
PBEsol~\cite{perdew_restoring_2008}.

\begin{figure}[t]
\includegraphics[width=1\columnwidth,angle=0]{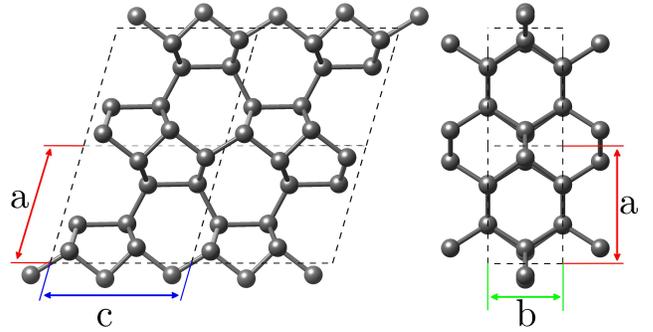} 
\caption{The structure of $M10$-carbon from two different angles. The left panel shows the 5- and 7-membered rings, while the right panel reveals the 6-membered rings.}
\label{fig:structure}
\end{figure}

Several MHM simulations were carried out with cells containing up to 8 atoms at 15~GPa within an unconstrained and thorough structural search, starting from different input configurations. We identified a novel, monoclinic carbon phase with $P2/m$ symmetry, which we call $M10$-carbon. It consists solely of $sp^3$ bonds and contains 8 atoms per cell. At ambient pressure, the unit cell parameters are given by $a=4.080$~\AA, $b=2.498$~\AA, $a=4.728$~\AA, $\alpha=\gamma=90^\circ$ and $\beta=73.96^\circ$. Two carbon atoms each occupy the crystallographic $2n$ sites at $(-0.1,0.5,-0.113)$ and  $(-0.132,0.5,0.421)$, and the $2m$ sites at $(-0.333,0,-0.466)$ and $(-0.325,0,-0.117)$. The overall structure is closely related to $M$-carbon, also consisting of 5- and 7-membered rings along the $b$-axis, while 6-membered rings are formed along the $c$-axis. In contrast to $M$-carbon, the 5-rings share the long edge, whereas in $M10$-carbon they share the short edge. The structure is illustrated in Fig.\ref{fig:structure}.

\begin{figure}[t!]  
\setlength{\unitlength}{1cm}
\includegraphics[width=1\columnwidth,angle=0]{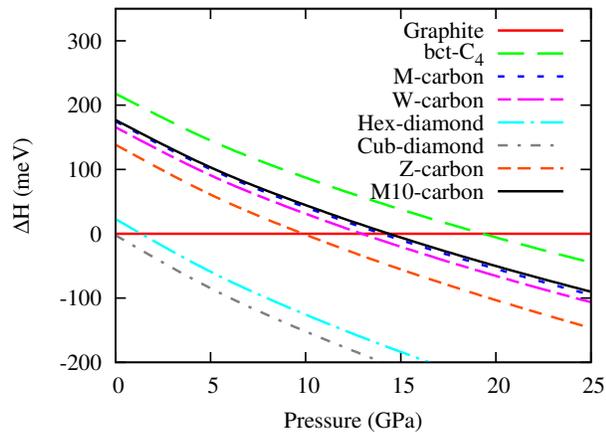}
\caption{Enthalpy differences per atom with respect to graphite of several carbon allotropes are shown as a function of pressure. The most promising candidate is $Z$-carbon since it crosses the graphite line at roughly 10~GPa. However, $M10$-carbon becomes thermodynamically accessible above 14.4~GPa .}
\label{fig:enthalpy}
\end{figure}

The dynamical stability was investigated by analyzing the phonon dispersion within the whole Brillouin zone. The density-functional perturbation theory~\cite{DFPT_S.Baroni} as implemented in {\sc abinit} was employed with a 12x12x12 $k$-point sampling and a 4x4x4 $q$-point mesh. No imaginary phonon frequencies were found, confirming the lattice stability of the phase. The thermodynamical stability of $M10$-carbon was investigated by computing its enthalpy within a wide pressure range. In Fig.~\ref{fig:enthalpy} the enthalpies of all proposed candidates for cold graphite are plotted with respect to graphite as a function of pressure. We neglected the zero-point vibrational energies in our calculations. $M10$-carbon becomes enthalpically favorable over graphite above a pressure of 14.4~GPa. As expected from the structural similarities with $M$-carbon, both $M10$- and $M$-carbon are very close in enthalpy throughout the whole pressure range.

In Table~\ref{tab:properties} we compare the structural properties of $M10$-carbon with other carbon allotropes. The bulk moduli $B_0$ were computed using the Murnaghan equation, and the Vicker's hardnesses $H_v$ were estimated with the method of Gao~\etal~\cite{gao_hardness_2003}. Like all of the investigated structures, $M10$-carbon is nearly as hard as diamond and has a very high bulk modulus, which could well account for ring cracks in diamond anvil cells~\cite{mao_bonding_2003}. Furthermore, an analysis of the electronic bandstructure was carried out, showing that $M10$-carbon is a wide band-gap semiconductor with an indirect DFT gap of 4.4~eV at 0~GPa.

\begin{figure}[t!]         
\setlength{\unitlength}{1cm}
\includegraphics[width=1\columnwidth,angle=0]{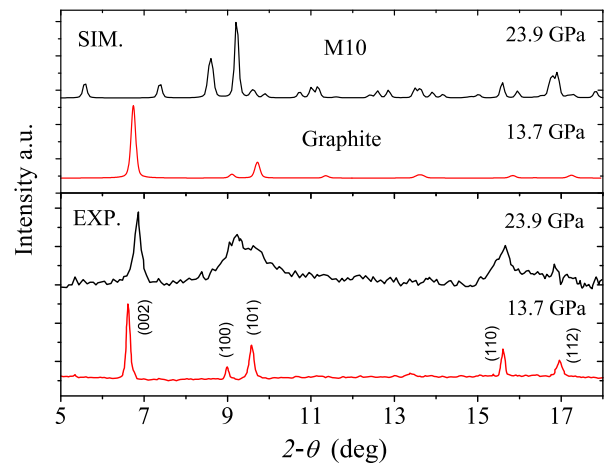}
\caption{The simulated XRD patterns are compared with the experimental results from Ref.~\cite{mao_bonding_2003} of cold compressed graphite. While at 13.7~GPa graphite was used to compute the simulated XRD pattern, $M10$-carbon was used for the simulation at 23.9\,GPa.}
\label{fig:xrd}
\end{figure}

Finally, we compare the XRD pattern of $M10$-carbon to experimental measurements from Ref.~\cite{mao_bonding_2003}, as illustrated in Fig.~\ref{fig:xrd}. The simulated pattern gives a good match to the experimental spectrum and could well explain the observed changes. It can therefore be expected that $M10$-carbon is present in samples of cold-compressed graphite above 14.4~GPa. However, other carbon allotropes show a similarly good agreement and thus the XRD pattern alone is by no way a conclusive evidence.

\begin{table}[b]

  \caption{Calculated and experimental data (where available) of the bulk moduli $B_0$ (in GPa), Vickers Hardness $H_v$ (in GPa) and volumes per atom $V_0$  (in \AA/atom) at 0~GPa for bct-C$_4$, $M$-, $W$-, $Z$-, $M10$-carbon and diamond.}
\begin{ruledtabular}
\begin{tabular} {l c c c c }
 Structure & Method   & $B_0$(GPa)& $H_v$(GPa) & $V_0$(\AA)  \\
    \hline
bct-C$_4$      &this work &428.2   & 93.5 & 5.82     \\
            &LDA~\cite{wang_low-temperature_2011} &433.7   &   & 5.83  \\

$M$-carbon  &this work &428.4   & 93.9 & 5.77       \\
            &  LDA~\cite{li_superhard_2009} & 431.2  & 83.1\ & 5.78   \\
$W$-carbon  &this work &427.5   & 94.2 & 5.75       \\
            &LDA~\cite{wang_low-temperature_2011} &444.5   &  & 5.76   \\ 
Cub-Diamond &this work &463.0   & 97.8 & 5.51       \\
            &LDA~\cite{wang_low-temperature_2011}  &466.3   &  & 5.52  \\
            &Expt.     &446\footnotemark[1]    & 60-120\footnotemark[2] & 5.67\footnotemark[1]  \\ 
$Z$-carbon  &LDA~\cite{amsler_crystal_2012} &441.5   & 95.4 & 5.66     \\
$M10$-carbon & this work &423.7 & 93.5&5.79
\end{tabular}
\end{ruledtabular}
\label{tab:properties}
\footnotetext[1]{Reference~\cite{occelli_properties_2003}}
\footnotetext[2]{Reference~\cite{brazhkin_our_2004}}

\end{table}

In conclusion we present a novel carbon allotrope that we call $M10$-carbon. It is a transparent, super-hard material which becomes enthalpically favorable over graphite at pressures above 14.4~GPa. Both the structural and enthalpical properties are very similar to the previously proposed $M$-carbon~\cite{li_superhard_2009}. The XRD pattern is in good agreement with experimental measurements, and although other carbon allotropes are enthalpically preferred at lower pressures, it could well be synthesized in samples of cold compressed graphite.

\acknowledgments Financial support provided by
the Swiss National Science Foundation is gratefully acknowledged.
JAFL acknowledges the CONACyT-Mexico. SB acknowledges support from
EU’s 7th Framework Programme (e-I3 contract ETSF) and MALM from the
French ANR (ANR-08-CEXC8-008-01). Computational resources were
provided by the Swiss National Supercomputing Center (CSCS) in Manno
and IDRIS-GENCI (project x2011096017) in France.

%

\end{document}